Results of Search for Magnetized Quark-Nugget Dark Matter from Radial Impacts on Earth

J. Pace VanDevender[1*], Robert G. Schmitt[2], Niall McGinley[3], David G. Duggan[4], Seamus McGinty[5], Aaron P. VanDevender[1], Peter Wilson[6], Deborah Dixon[7], Helen Girard[1], and Jacquelyn McRae[1]

1. VanDevender Enterprises LLC, 7604 Lamplighter Lane NE, Albuquerque, NM 87109 USA; pace@vandevender.com
2. Sandia National Laboratories, Albuquerque, NM 87185−0840, USA; pace@vandevender.com
3. Ardaturr, Churchill PO, Letterkenny, Co. Donegal, Ireland; niallmacfhionnghaile@gmail.com
4. National Park and Wildlife Service, Glenveagh National Park, Church Hill, County Donegal, F92XK02, Ireland; dave.duggan@chg.gov.ie
5. Church Hill, Letterkenny, County Donegal, F928982, Ireland, pace@vandevender.com
6. School of Geography and Environmental Sciences, Ulster University, Cromore Road, Coleraine, Co. Londonderry, BT52 1SA Northern Ireland, UK; p.wilson@ulster.ac.uk
7. DKD Engineering Inc., 801 El Alhambra Cir. NW, Los Ranchos, NM 87107, USA; ddixon@dkdengineeringinc.com

*Correspondence: pace@vandevender.com

**Abstract:** Magnetized quark nuggets (MQNs) are a recently proposed dark-matter candidate consistent with the Standard Model and with Tatsumi's theory of quark-nugget cores in magnetars. Previous publications have covered their formation in the early universe, aggregation into a broad mass distribution before they can decay by the weak force, interaction with normal matter through their magnetopause, and a first observation consistent MQNs: a nearly tangential impact limiting their surface-magnetic-field parameter $B_o$ from Tatsumi's $\sim 10^{12+/-1}$ T to $1.65 \times 10^{12}$ T +/- 21%. The MQN mass distribution and interaction cross section strongly depend on $B_o$. Their magnetopause is much larger than their geometric dimensions and can cause sufficient energy deposition to form non-meteorite craters, which are reported approximately annually. We report computer simulations of the MQN energy deposition in water-saturated peat, soft sediments, and granite, and report the results from excavating such a crater. Five points of agreement between observations and hydrodynamic simulations of an MQN impact support this second observation being consistent with MQN dark matter and suggest a method for qualifying additional MQN events. The results also redundantly constrain $B_o$ to $\geqslant 4 \times 10^{11}$ T.

1. Introduction

We report results of computer simulations and observations from field work that indicate that at least one non-meteorite impact crater was formed by an impactor with mass density comparable to nuclear density, with mass ~5 kg and with energy deposition of ~ 80 MJ/m. We show that these results are consistent with ferromagnetic Magnetized Quark Nuggets (MQNs), which are a relatively new candidate for dark matter. These observations from non-meteorite craters may also be consistent with some other dark-matter candidates. Non-meteorite craters are reported in the popular press approximately once per year and they may offer an opportunity to test hypotheses for dark matter.

This is the fifth paper on MQNs. For your convenience, the introduction will summarize the basic characteristics of MQNs that were demonstrated in previous papers and place MQNs in the context of current research on nuclearites, which are non-magnetic quark nuggets.

*1.1. From Dark Matter to Quark Nuggets, Nuclearites, and MQNs*

Dark matter [1–6] comprises approximately 85% of the mass in the universe. After decades of searches for experimental and observational evidence supporting any of many candidates for dark matter, the nature of dark matter is still a mystery [6–7]. A quark nugget is a dark-matter candidate that is composed of up, down, and strange quarks [8–15]. Quarks are constituents of many particles in the Standard Model of Particle Physics [16]. Quark nuggets are a Standard-Model candidate for dark matter that has not been excluded by observations [13,14,15]. A current summary of quark-nugget research can be found in reference [17].

Many physicists know of Witten's [8] 1984 proposal that a Quantum Chromodynamics (QCD) phase transition would have permitted stable quark nuggets to have formed in the early universe and function as dark matter. They also know that the QCD phase transition is not currently supported. Consequently, some physicists concluded quark nuggets are no longer viable. However, Aoki et al. [18] showed that an analytic crossover process would have formed quark nuggets in the early Universe without a phase transition. Recent simulations conducted by T. Bhattacharya et al. [19] support the crossover process. Both of these papers and others assert that quark nuggets could be formed without the phase transition and they are still a viable candidate for dark matter.

The search for quark-nugget dark matter continues primarily as the search for nuclearites, which are quark nuggets with an interaction cross section equal to their geometric cross section. Burdin, et al. [14] included nuclearites in his review of non-collider searches for stable massive particles in 2015. The work he reviewed analyzed data to find the upper limit to the nuclearite flux while assuming that 100% of the local dark-matter density is composed of nuclearites of a single mass. They concluded that the combined results of all the searches do not exclude nuclearites from 0.1 to 10 kg, $10^6$ to $10^{14}$ kg and $10^{19}$ to $10^{21}$ kg with the single-mass model. The Joint Experiment Missions for Extreme Universe Space Observatory (JEM-EUSO) is the next big step in looking for nuclearites, which will feature a 2.5-m telescope with a wide (60⁰) field of view operating from the International Space Station. In addition to its other missions, JEM-EUSO could close the 0.1-to-10 kg gap in the single-mass nuclearite model with just 1 to 100 days of data for 0.1-kg to 10-kg nuclearites, respectively [20]. All of these results assume that 100% of the dark matter density is composed of nuclearites of a single (or average) mass and their interaction cross section is their geometric cross section at nuclear mass density.

These reviews and plans did not consider MQNs, which were first published in 2017 [21] after twenty years of research to explain the anomalies that we now associate with non-meteorite impacts, the subject of this paper. MQNs differ from nuclearites in that: 1) MQNs are ferromagnetic, as explained in the next subsection, 2) are predicted to have a broad mass distribution between ~ $10^{-24}$ kg and ~$10^{+6}$ kg [15], as illustrated in Figure 1 instead of single-mass nuclearites, and 3) interact with normal matter through their magnetopause [21] which may be millions of times larger than their geometric cross section, as quantified by equation (1). The large mass distribution means that capabilities like JEM-EUSO would require ~9 years of continuous and dedicated observations to test the MQN hypothesis, as discussed in Section 1.5, below. The much larger interaction cross section per unit mass means that lower-mass and more abundant MQNs do not reach deeply buried ancient mica or space-based track recorders behind spacecraft walls that are discussed in reference [14] or the scintillators in underground observatories [22]. The enhanced cross section also means that the energy of larger MQNs impacting Earth or the Moon is deposited in many kilometers instead of passing through the body [14], as usually assumed; the extremely high energy deposition excites strongly attenuated shear modes in rock that complicate seismic detection, in contrast to the elastic modes assumed. The broad mass distribution and the large interaction cross section both arise from MQN ferromagnetism.

*1.2. Theoretical Basis of Ferromagnetism in MQNs*

MQN ferromagnetism is based on the existence of magnetars, which are pulsars with magnetic field ~300 times the magnetic field of neutron stars. The much larger magnetic field implies a different physical nature for magnetars, such as a quark nugget core. Xu [23] has shown that the low electron density, as permitted in stable quark nuggets, limits surface magnetic fields from ordinary electron ferromagnetism to $\sim 2 \times 10^7$ T. Tatsumi [24] examined ferromagnetism from a One Gluon Exchange interaction in quark nuggets and concluded that the surface magnetic field could be $\sim 10^{12 +/- 1}$ T, which is sufficient for explaining the magnetic field inferred for magnetar cores. The result needs to be confirmed with relevant observations and/or advances in QCD calculations because the result depends on the currently unknown value of the QCD coupling strength [16] $\alpha_c$ at the ~90 MeV energy scale of the strange quark.

We are exploring the implications of such Magnetized Quark Nuggets (MQNs) to explain the anomaly of non-meteorite impacts that started our investigations and the anomaly of dark matter because magnetars exist [25] with a magnetic field that is ~300 times that of neutron-star pulsars [26] and since such a large magnetic field is perhaps uniquely consistent with ferromagnetic quark nuggets.

*1.3. From Ferromagnetism to MQN Stability and Mass Distribution*

The previously published theoretical results [15] show that MQNs would have originated at time t ~ 65 μs when the Universe had a temperature of ~ 100 MeV according to the Standard Model of Cosmology [2]. At that temperature, $\Lambda^0$ particles (consisting of one up, one down, and one strange quark) could form [27]. The simulations of their aggregation as a ferromagnetic liquid under the long-range magnetic force, similar to simulations of inelastic collisions of particles in nucleosynthesis under the short-range nuclear force, showed that MQNs magnetically aggregate [15] into a broad mass distribution of stable ferromagnetic MQNs before they could decay. In the extremely high mass density of the early Universe, aggregation happened with an initial time scale of ~1.5 ps, so aggregation dominated the ~10 ps decay by the weak interaction. Because current day accelerator experiments have a much lower mass density, aggregation does not compete with decay, so stable MQNs are not formed in those experiments.

After time t ≈ 66 μs after the big bang, the simulated mean of the MQN mass distribution is between $\sim 10^{-6}$ kg and $\sim 10^4$ kg, depending on the surface magnetic field $B_o$. The corresponding mass distribution is sufficient for MQNs to meet the requirements [13,15] of dark matter in the subsequent processes, including those that determine the Large Scale Structure (LSS) of the Universe and the Cosmic Microwave Background (CMB).

Throughout this paper, we will use $B_o$ as a key parameter that defines the mass distribution of MQNs. The value of $B_o$ equals Tatsumi's surface magnetic field $B_S$ if the mass density of MQNs $\rho_{QN} = 10^{18}$ kg/m$^3$ and density of dark matter was $\rho_{DM} = 1.6 \times 10^8$ kg/m$^3$ when the temperature of the universe was ~100 MeV [15,27]. If better values of $\rho_{QN}$ and $\rho_{DM}$ are found, then the corresponding values of $B_S$ can be calculated by multiplying the $B_o$ from our results by $(1 \times 10^{-18} \rho_{QN})(6.25 \times 10^{-9} \rho_{DM})$.

The previously published MQN papers, especially the observational paper [17], narrowed the allowed range of $B_o$ from Tatsumi's ~$10^{12+/-1}$ T to $1.65 \times 10^{12}$ T +/- 21% to be consistent with the maximum mass of MQNs that is allowed for a given value of $B_o$. The integrated MQN number flux for MQN mass $\geq M_{MQN}$ has been derived from the mass distribution assuming a constant velocity of 230 km/s, the velocity of the solar system through the galactic halo. The integrated flux is useful for evaluating event rates and it is shown in Figure 1 for the current range of $B_o$ that is consistent with observations.

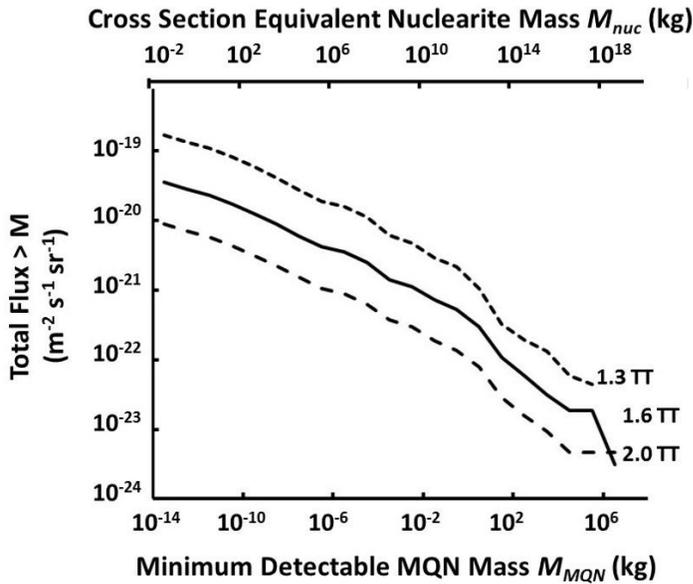

**Figure 1.** The integral of the MQN number density from minimum detectable MQN mass $M_{MQN}$ to infinity times 230 km/s, the velocity of the solar system through the galactic halo, gives the predicted number flux for mass $\geq M_{MQN}$ plotted on the x-axis for range of currently allowed values of $B_o$. For $B_o = 1.65 \times 10^{12}$ T and atmospheric mass density (~$10^{-3}$ kg/m$^3$) at the 50 km altitude to be probed for nuclearites by JEM-EUSO, the nuclearite mass $M_{nuc}$ with the same interaction cross section as $M_{MQN}$ is plotted above the graph.

*1.4. Comparison of MQN and Nuclearite Interaction Cross Sections*

Like Earth, MQNs have a dipole magnetic field. Additionally, like Earth, they also have a magnetopause, which interacts with inflowing plasma as Earth's magnetopause interacts with the solar wind. We assume the extremely high MQN velocity relative to the surrounding matter assures the plasma temperature is sufficient to fully ionize the inflowing matter. Under that assumption, Equation (1), derived from reference [21], gives the ratio of MQN cross section to its geometric cross section, and Equation (2) gives the nuclearite mass $M_{nuc}$ with the same interaction cross section as the MQN of mass $M_{MQN}$.

$$\frac{\sigma_{MQN}}{\sigma_{nuc}} \approx \left(\frac{2B_o^2}{\mu_0 \rho_p v^2}\right)^{\frac{1}{3}} \text{ and} \quad (1)$$

$$M_{nuc} \approx \left(\frac{2B_o^2}{\mu_0 \rho_p v^2}\right)^{\frac{1}{2}} M_{MQN}, \quad (2)$$

for $\rho_p$ = the mass density of surrounding plasma, $\mu_o$ = permeability of free space, $B_o$ = MQN surface magnetic field parameter, and $v$ = velocity of MQN relative to the surrounding plasma. The ratio of interaction cross sections in Equation (1) and equivalent nuclearite mass in Equation (2) depend on the mass density of surrounding material. The top scale shown in Figure 1 shows the equivalent nuclearite mass for JEM-EUSO observations at the appropriate altitude.

*1.5. Estimated Observation Time for JEM-EUSO to Test MQN Hypothesis*

Reference [20] indicates that only 24 hours of observation will be needed for JEM-EUSO to determine whether nuclearites of single-mass $M_{nuc} = 10^{26}$ GeV/c$^2$ = 0.16 kg have a flux consistent with the Galactic dark-matter limit. Equation (2) gives $M_{nuc} = 1.8 \times 10^{12} M_{MQN}$ for the same cross section and for $\rho_p = 10^{-3}$ kg/m$^3$ appropriate for the 50 km altitude to be observed by JEM-EUSO. Conversely, the MQN mass that is equivalent to a nuclearite with $M_{nuc} = 0.16$ kg nuclearite mass is $M_{MQN}$ = ~$10^{-13}$ kg. Therefore, JEM-EUSO will be able to make the same judgement regarding MQNs with $M_{MQN} > 10^{-13}$ kg if the acceptance for MQN mass > $M_{MQN}$ is the same as it is for $M_{nuc} = 0.16$ kg. However, the required time scales vary inversely with flux, which is $10^{-16}$ m$^{-2}$ s$^{-1}$ sr$^{-1}$ for single-mass nuclearites and $3 \times 10^{-20}$ m$^{-2}$ s$^{-1}$ sr$^{-1}$ for distributed mass MQNs with $B_o = 1.65 \times 10^{12}$ T. Therefore, testing the MQN hypothesis with JEM-EUSO would take ~ 3,000 days or ~9 years of dedicated observation time. Because JEM-EUSO has many demands for its observation time, it is impractical.

*1.6. Definition of Non-Meteorite Craters and their Utility for Testing Dark-Matter Hypotheses*

This paper investigates the anomaly of non-meteorite craters as a means to test the MQN hypothesis and explore other dark-matter candidates. Non-meteorite craters are defined as craters formed:

1) without an observable luminous streak,

2) without breakup in an air shower [28] and dispersal, so no crater is formed,

3) without meteorite material found in and/or around the impact site, and

4) without evidence of causation by human or other natural causes (e.g. the energetic release of methane from global warming), but

5) with sufficient energy deposition to form an impact crater.

Criteria 2 through 4 are straightforward, but criterion 1 requires some explanation, since energy is being deposited in the atmosphere by a hypervelocity object. MQNs interact with surrounding matter [21] through their magnetopause, which is the boundary between their compressed magnetic field and the plasma pressure from inflowing matter. Their magnetopause is much larger than their nuclear-density core, but it is still quite small. For example, a 5-kg MQN moving through air at sea level with speed of 230 km/s has a nuclear density core with radius of ~ $7.5 \times 10^{-10}$ m and a magnetopause radius of $1.5 \times 10^{-6}$ m. At a distance of 15 km, which is typical for a city, the apparent magnitude [10] of the corresponding luminosity at a distance of 15 km would be -4.4, which is approximately that of Venus on a clear night. The characteristic radius of the shock wave and characteristic cooling time of the shock temperature [29] are ~$8 \times 10^{-4}$ m and ~$2 \times 10^{-6}$ s, respectively. Even after expansion and cooling for a hundred characteristic times, the angular diameter at 15 km is only 2.2 arc seconds, which is only ~3.3% of the angular diameter of Venus at closest approach. In addition, the transit through the 8-km e-folding distance of the atmospheric density is only 0.034 s. It takes 0.25 s for humans to perceive an unexpected object in their field of view as a thing [30], so, even if it could be seen, it would not be recognized before it had gone. Therefore, non-meteorite craters are not associated with human-observable events.

This paper focuses on an observational test of the theory, not on advancing the theory of MQNs. We simulate the interaction of an MQN with a geophysical three-layer witness plate, and then test the resulting signature of an MQN impact with observations from one non-meteorite impact crater. The simulations connect the previously published theory to the observations at the crater for a test of the MQN hypothesis.

We use the terms meteors and meteorites to refer to bodies that are composed of normal matter, i.e. atoms that are held together by the electromagnetic force. The material strength that is associated with the electromagnetic force is weak. Thus, meteors and meteorites (as defined in this paper) must be quite large to survive intact and they do not make small craters. Nuclear-density quark nuggets are held together by the strong-nuclear force, so all of them survive passage through the atmosphere.

Craters that show no evidence of meteorite impact are reported in the press approximately once per year. Therefore, the event rate for non-meteorite craters is sufficient to allow the phenomenon to be studied, if access to the craters can be obtained. Three events in three years have been recently reported.

1. ~12-m diameter crater near Managua, Nicaragua, on September 6, 2014 [31].

2. ~1-cm diameter crater in Rhode Island, USA, on July 4, 2015 [32].

3. ~60-cm diameter crater in Tamil Nadu, India, on February 6, 2016 [33].

None of these impacts was preceded by a luminous track in the sky, no meteorite material was found in or near any of the craters, and experts who reviewed the news reports concluded that these events were not meteorite impacts, as reported in references [31,32,33]. In the absence of a scientific basis for impacts that form craters without luminous tracks and without meteorite fragments, these experts attributed them to human-caused explosions by default. Our results indicate that MQNs can also cause non-meteorite craters and they should be considered in future investigations.

Each non-meteorite event provides a large-target opportunity to test the MQN dark-matter hypothesis. Because a multi-layer witness plate provides more information than a single-layer one, peat bogs on top of soft sediments and bedrock offer particularly useful opportunities. Sections 3.1 through 3.2 report the results of hydrodynamic simulations and analyses of MQNs interacting with a three-layer witness plate of peat, clay-sand mixture, and granite bedrock in County Donegal, Ireland. Other peat bogs could also provide such opportunities. However, County Donegal has the advantages of: 1) a granite bedrock within excavation range of the surface, 2) a friendly and supportive population, 3) a governing authority over peat bogs that can grant permits for exploration, and 4) maximum exposure to the directed flux of dark matter. The last advantage arises because dark matter streams into Earth as the solar system moves around the galactic center and through the dark-matter halo. That direction of motion is right ascension 18h 36m 56.33635s, declination $+38° \ 47′ \ 01.2802″$ and it is always above the horizon for latitudes that are greater than that declination, including the latitude of County Donegal.

*1.7. Significance of this Paper*

MQN interaction provides at least three measurable signatures of MQN dark matter:

1) hypervelocity (which generally refers to velocities > 3,000 m/s at which the material strength is much less than internal stresses) atmospheric transit without luminous streak and without breakup in an air shower [28], but with energetic (> 1 kJ/m) energy deposition and with multi-meter transit through solid-density matter,

2) electromagnetic emissions (kHz to GHz) from the rotating magnetic dipole after transit through matter [34], and

3) magnetic levitation of rotating MQN magnetic dipole after transit through matter [17] by induced currents in adjacent conducting material or magnetic levitation of static magnetic dipole above a superconductor.

A systematic attempt to detect MQNs through the first signature was attempted [21] by looking for acoustic signals from MQN impacts in the Great Salt Lake in Utah, USA. Even though the first method monitored ~30 sq-km (i.e. ~30 times the cross section of the IceCube Neutrino Observatory at the South Pole) for impacts with MQN mass $\geq 10^{-4}$ kg, none were detected [15]. The null results implied that analysis that is based on average mass, which is the usual assumption for dark-matter candidates, may be inadequate and motivated the detailed computation of the MQN mass distribution. Those results show a much larger detector and/or much longer observation times are required.

Reference [34] describes how MQNs passing through matter spin up to MHz frequencies and emit radiofrequency energy (the basis of the second signature), and proposes using Earth's magnetosphere as a sufficiently large detector to obtain enough events. The large uncertainty in $B_o$ from Tatsumi's theory is a major impediment to designing and fielding such an experiment.

Reference [17] describes an extremely rare episodic observation and supporting simulations of the third method. That method is too rare to provide enough data to measure the mass distribution and it provides sufficient statistics for discovery of MQNs. However, those results do limit the key parameter $B_o$ to $1.3 \times 10^{12}$ T $\leq B_o \leq 2 \times 10^{12}$ T and they permit the design of the systematic experiment based on the second signature with Earth's magnetosphere as the detector area.

Such a project requires major investment. Additional episodic data would help to justify the project to obtain systematic data. The current paper provides additional episodic observational and supporting computational results of the first method, based on non-meteorite crater impacts on Earth. We also discuss extending the results to find a statistically significant number of MQN impacts.

*1.8. Organization of the Paper*

Sections 3.4 through 3.8 report the observations and associated analyses from excavating a non-meteorite impact that occurred in County Donegal in May of 1985. We find the crater and subsurface damage to be consistent with an MQN impact that deposited ~ 80 MJ/m in the water-saturated peat. The corresponding MQN mass depends on the value of $B_o$. MQN mass distributions that are consistent with ~80 MJ/m energy deposition provide an additional exclusion of $B_o < 4 \times 10^{11}$ T. For the most likely range of $1.3 \times 10^{12}$ T $\leq B_o \leq 2 \times 10^{12}$ T, ~80 MJ/m energy deposition corresponds to MQN mass of 5 +/- 1 kg.

Section 4 discusses the results, alternative explanations, and a potential method for more efficiently determining whether a candidate impact is consistent with a deeply penetrating MQN impact or with some surface phenomenon. In principle, such a method could locate additional MQN events.

*1.9. Declaration of Controversial Topic*

The editorial policy for this journal requires authors to declare whether an article is controversial. For the last four decades, searches for dark-matter candidates have focused on particles that are Beyond the Standard Model (BSM) of Particle Physics. Because MQNs are composed of Standard Model quarks and the theory for the ferromagnetic state of MQNs is an approximate solution within the Standard Model, the MQN hypothesis for dark matter does not require a BSM particle and it may be considered controversial by the BSM community. However, a potential solution to dark matter for the Standard Model of Cosmology and not requiring a hypothetical BSM particle may be less controversial to others because it is a modest extension of well-established physics.

*1.10. Limitations of these Results*

We report the results from just one episodic event, which we calculate in Section 4.2 to have a < 2% of arising randomly from unknown effects. Nevertheless, we conclude that many more instances are required to determine whether or not MQNs exist and contribute to dark matter. The event that is presented in this paper suggests a method for adding more instances of MQN interaction by qualifying other non-meteorite impacts, which occur approximately annually.

## 2. Materials and Methods

This study used

1. computational 2D and 3D hydrodynamic simulations, as described in the results section and in movies of pressure, mass density, and temperature in supplementary dataset at https://doi.org/10.5061/dryad.cc2fqz641 and associated analyses of energy deposition in the multi-layers of peat bog in the results section;

2. original field work at the location of a non-meteorite impact in May of 1985 in County Donegal, Ireland, and associated analyses in the results section;

3. additional details in Appendix A: Excavations, to assist an independent team to extend our findings; and,

4. potential sites in Appendix B: coordinates and description of deformations in peat-bog survey, for future investigations if suitable sensor technology can be developed.

## 3. Results

*3.1. Hydrodynamic Simulation of MQN Impact in Three-Layer Witness Plate*

Two- and three-dimensional simulations with the CTH hydrodynamics simulation software [35,36] were conducted to investigate MQN interactions with a three-layer witness plate of peat-bog, clay-sand mixture, and granite bedrock. Two-dimensional simulations examined the craters that formed by MQN impacts as a function of MQN mass and the $B_o$ parameter. Three dimensional simulations investigated the circularity of the crater as a function of angle relative to vertical and guided the excavation of an impact site in County Donegal, Ireland.

In both cases, the MQN deposits energy along its path and within the magnetopause radius [21] (e.g. ~1 mm radius for a 30 MJ/m energy deposition from a 1-kg MQN for $B_o \approx 10^{12}$ T). The interaction produces a plasma with an initial temperature of ~2,100 eV. We did not have access to any computer program that could reliably resolve the dynamics of the initial channel (requiring <<1-ns and ~100-µm resolution) and still simulate acoustic propagation in two dimensions over many meters and for many milliseconds. Other studies have shown that turbulent mixing with colder material dominates the early dynamics of the interaction and produces a channel of ~ 1 eV temperature within a radius that preserves the energy per unit length. Therefore, we approximated the post-turbulence phase of the plasma channel as a cylinder with the mass density of the peat, clay-sand, or granite. The temperature was varied from 0.5 to 1.55 eV and the radius was varied to give the specified energy per unit length. The equation of state was imported from SESAME4 [36] data. The results were essentially independent of temperature over that range and they validated the assumption that energy/length is the dominant variable.

The fluid above the peat was atmosphere at standard temperature and pressure. The simulated depth of the peat was the actual 0.7 m of the Irish peat bog with an initial density of $1.12 \times 10^3$ kg/m$^3$ and sound speed of $1.46 \times 10^3$ m/s. The 4.7 m-thick clay-sand layer was simulated with a 1.0 meter thick layer and with an initial density of $2.02 \times 10^3$ kg/m$^3$ and sound speed of $2.2 \times 10^3$ m/s. The granite layer was simulated with a 0.3-m layer with initial density of $2.6 \times 10^3$ kg/m$^3$ and sound speed of $5.0 \times 10^3$ m/s. The bottom of each simulation was unmovable, and material could freely exit from the other boundaries.

Two-dimensional hydrodynamic simulations were conducted with 1, 3, 9, 27, 81, and 243 MJ/m energy deposition. They show that a shock wave reflects off the mass discontinuities and it propagates radially outward in all three layers. Low-density and high-temperature material in the central channel is ejected into the atmosphere. Lower temperature material behind the shock wave moves radially and almost one-dimensionally outward. Finally, the peat distorts two-dimensionally in response to the velocity field that it has acquired and the shear planes that have developed within the peat.

Figure 2 shows representative results of the density maps for times when the material velocities are well below their peak values.

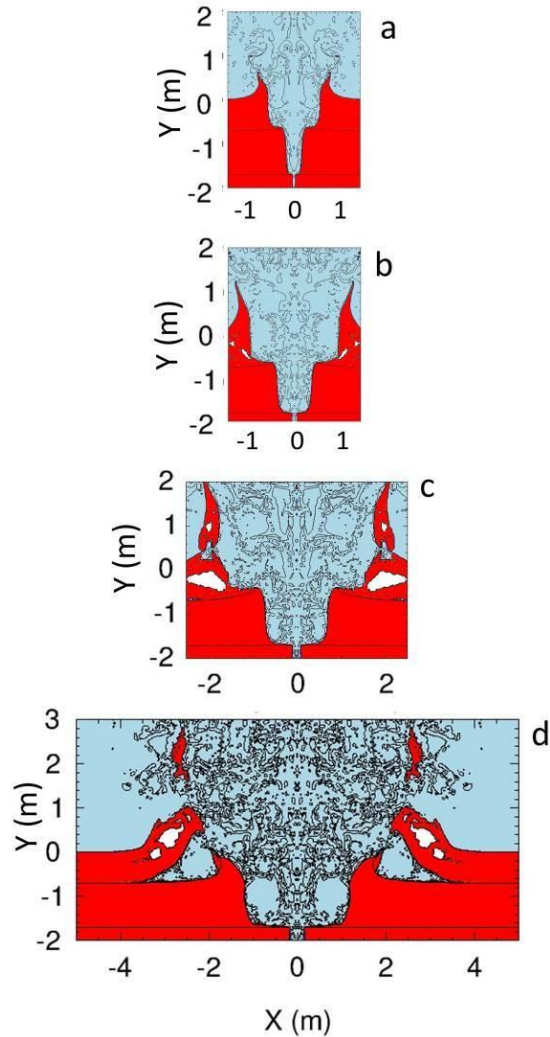

**Figure 2.** Representative density maps are shown for times when radial expansion of the channel in the clay-sand layer is approaching its maximum: **a)** 9 MJ/m at t = 20 ms, **b)** 27 MJ/m at t = 25 ms, **c)** 81 MJ/m at t = 40 ms, and **d)** 243 MJ/m at t = 55 ms. The red material represents three layers separated by black lines. From the bottom up, the layers are 0.3 m of granite, 1.0 m of clay-sand, and 0.7 m of peat with initial density of $1.12 \times 10^3$ kg/m$^3$. The blue area is atmospheric air. The white spaces are voids at shear planes. Movies of pressure, density, and temperature are available online (accessed on 14 2 2021: https://datadryad.org/stash/share/Lt7dMvxEAUWNnkfKt2xPg2l5TuWz7Bbec67iY4Kvazg.

The 81 MJ/m case shown in Figure 2c is especially relevant to the crater discussed below. The shear planes and voids form relatively smooth sides of the peat crater. Most of the peat is ejected in small fragments into the atmosphere. Larger pieces of peat are shown as vertical pieces about to be ejected radially away from the crater. The channel is almost one-dimensional in the clay-sand and granite.

Figure 3 shows the summary results of simulations for 1 MJ/m to 243 MJ/m.

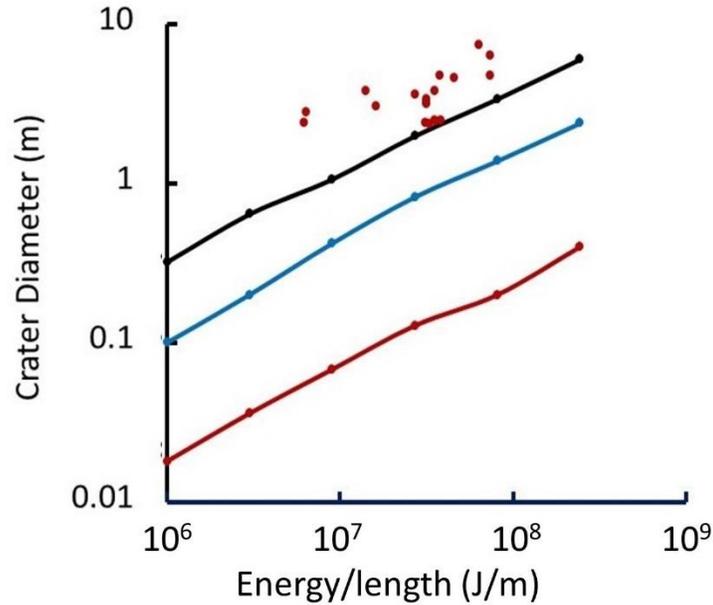

**Figure 3.** Solid lines show crater diameter in granite (red), clay-sand (blue), and peat (black) as a function of the energy/length from CTH simulations. The data points show the diameter of fractured granite from reference [37].

The central channel in the granite that is shown in Figure 2 is caused by the compressive pulse that decreases rapidly with increasing distance from the high-energy-density center. When the compressive pulse reflects at the clay-sand boundary, it becomes a tensile pulse and breaks the rock in tension. Because the tensile strength of granite is only ~ 1.5% of the compressive strength [38], the diameter of fractured granite is much larger than the diameter of the compressed channel [37]. The fracture diameter depends on the geometry [38] and composition [37,38] of the explosive, shock impedances of both materials, and distance [37] to material with lower shock impedance. However, those effects are secondary to the main trend, as shown by the scatter in fracture data presented in Figure 3. Over a wide range of parameters, the fracture diameter from the tensile strain is approximately a factor of 30 larger than the diameter of the channel that is caused by compressive strain.

*3.2. Potential for Liquefaction and Flow of the Clay-Sand Layer*
Peat is ejected into the atmosphere and it leaves a crater with smooth sides that formed from the shear planes. The granite layer is fractured around the path of an MQN; however, the channel in the fully water-saturated clay-sand of the County Donegal peat bog is very likely to undergo liquefaction and close the channel within tens of seconds after the passage of an MQN. Fine-grained soils, i.e. silts and clays for which the percent of dried soil passing through a No. 200 standard sieve (i.e. 0.074-mm diameter openings) exceeds 50%, require careful testing and analysis to determine whether or not they will undergo liquefaction under an impulse or shaking, according to Boulanger and Idriss [39]. Conversely, soils with much less than 50% passing through a No. 200 sieve are much more likely to liquefy when they are saturated with water. We analyzed the clay-sand layer in the County Donegal peat bog and found that it is composed of ~10% rock of ~1 cm diameter, ~20% soil that does not pass a 1 mm screen, and 11% soil passing through a No 200 sieve. Therefore, the fine-grained portion is only ~11% of total mass or, more conservatively, 19% of the sub-mm sized content and well within the < 50% criterion for susceptibility to liquefaction.

Owen and Moretti [40] identified five conditions that contribute to liquefaction-induced soft-sediment deformation in sands under a transient increase in pore fluid pressure: 1) fine to medium-sized grains of sand, 2) high porosity, 3) high percent saturation with water, 4) low overburden pressure (<10 m of overburden), and 5) no previous liquefaction. The clay-sand layer between the peat and granite satisfies all five conditions. In addition, Owen and Moretti cite impact by extra-terrestrial objects as a likely trigger for liquefaction. Therefore, we conclude that the clay-sand layer is very likely to have undergone liquefaction and obscured the channel within tens of seconds after impact.

*3.3. Simulations on Circularity of MQN Crater as a Function of Entrance Angle*

In addition to identifying the signature of MQN impacts, CTH simulations examined the circularity of the crater in the peat bog as a function of entrance angle relative to vertical. The information is helpful in identifying the likely path of the MQN through the liquefied intermediate layer to the bedrock.

Simulations modeled channels at 0°, 15°, 30°, 45°, and 60° to the vertical and between the surface and an immovable solid at 1.0-m depth. The simulated MQN instantaneously deposited 30 MJ/m energy density, as described above. Because of the low strength of the peat, the crater continues to grow for an extended period of time. Each simulation was stopped at 10 ms in order to reasonably simulate the relative effects of the impact angle on the crater dynamics. Figure 4 shows representative profiles.

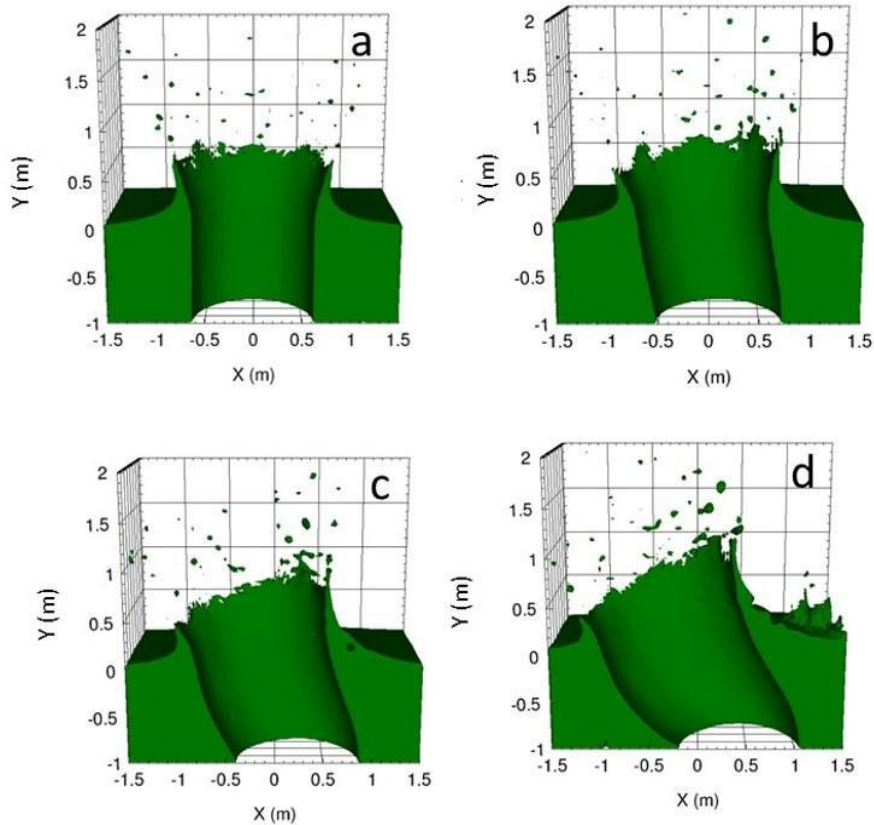

**Figure 4.** Mass density profiles of peat at $t = 10$ ms after the start of a simulated MQN interaction depositing 30 MJ/m on a trajectory inclined at a) 0°, b) 15°, c) 30°, and d) 45° from the vertical.

At $y = -0.5$ m, the ratio of major to minor axes is approximately $\cos^{-1}(\theta)$, as expected for a cylinder intersecting a plane at angle $\theta$. However, Figure 4 shows that the peat on the right-hand edges is forced against low-density air, while the peat on the opposite side is forced against higher-density peat. The less-impeded peat moves more. Therefore, the asymmetry is enhanced near the rim of the crater. Using the crater shape to estimate $\theta$ gives a maximum angle for the trajectory.

*3.4. Non-Meteorite crater in May 1985 near Glendowan, County Donegal, Ireland*

A non-meteorite impact occurred in the middle of May of 1985, on Stramore Upper, near Glendowan, County Donegal, Ireland at 54° 58.257′ N, 8° 0.408′ W. It was reported in the *Donegal People's Press*, May 31, 1985. The article said that it occurred "when people were walking their dog"; that would be about 18:00 hours GMT.

The site is on Common Land with rights assigned to a group of nearby landowners, who kindly allowed our research. The National Parks and Wildlife Service has authority over the land and it granted us a permit to excavate the site.

Two of us (D. D. and S. McG.) investigated the crater the day after the event, as part of our duties as Park Rangers. The inside sloped surface of the crater was very smooth. A few-centimeter diameter hole or "pointy" depression was present in the dirt at the center of the crater bottom. There was a distinct, ~2 cm high lip on the peat edge of the crater. Pieces of the bog were scattered up to 10 m away. A visual search of the crater and the surrounding area did not find any meteorite material. The crater filled with water within two days. The water prevented sub-surface searches and, to our knowledge, preserved the site until our team excavated it.

Figure 5 shows the relatively smooth sides, which are very unusual for craters that are produced by surface explosives in peat bogs. In addition, large pieces of peat were scattered approximately 10 m away. The smooth sides, diameter of the crater, and energetically-detached ejecta were consistent with the CTH simulation and they confirm that the simulation was accurately modeling the crater formation.

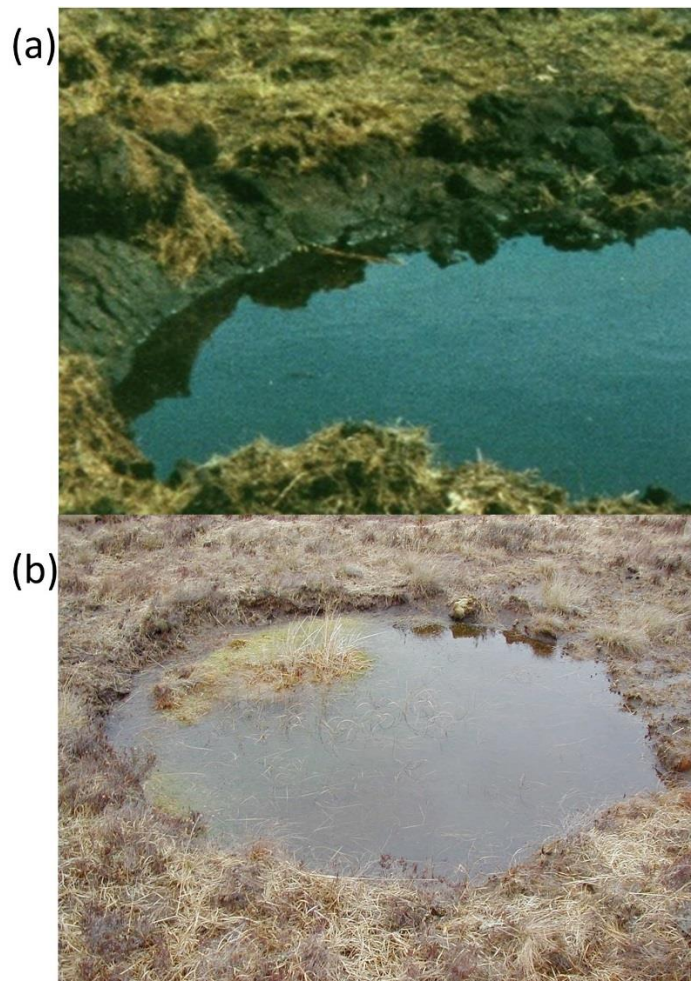

**Figure 5.** a) Photograph of the crater soon after the event in 1985 illustrate good agreement in the actual and simulated profile of crater sides. b) Photograph of the crater in March 2005 showing the full diameter and circularity before excavation.

The crater has a diameter of 3.984 ± 0.065 m in 2006. The yield strength of the peat was measured and found to be 530 ± 120 kN m$^{-2}$. Figure 2c and Figure 3 give an energy/meter of

~80 MJ/m for a 4.0-m diameter crater. Uncertainties in the equation of state variables imply a fidelity of +/- 20%.

The shape of the crater was measured in 2006, before it was distorted by investigations. The best fit to an ellipse gives a 1.030 ± 0.005 ratio of major to minor axes and it corresponds to $\theta \leq 15°$, as shown in Figure 4a or Figure 4b. The major axis aligned east-west. Therefore, the excavation was planned to explore the volume within 15° of vertical and optimized for the east or west of center.

*3.5. Excavations of the 1985 Non-Meteorite Crater in County Donegal, Ireland*

Field work a third of the way around the world and in a protected wilderness area is challenging at best. However, it is the least expensive way to test the MQN dark-matter hypothesis. The additional information presented in Appendix A should assist independent groups in learning from our experiences and re-excavating the site.

The site was excavated in three stages, as shown in Figure 6.

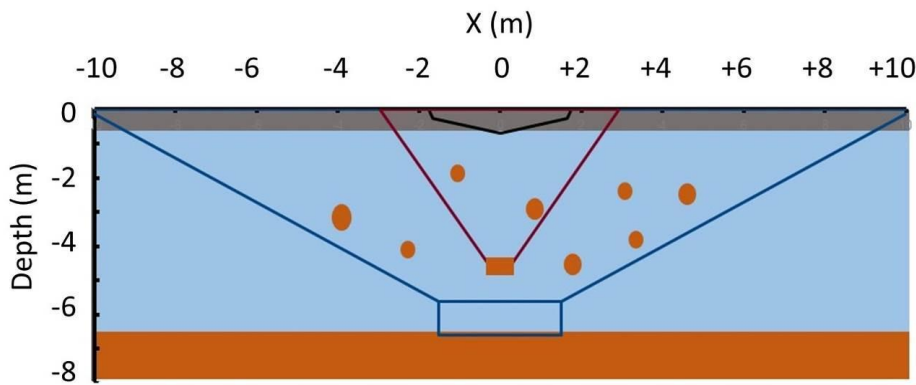

**Figure 6.** Cross-sectional view of the three excavations: 2017 (black line), 2018 (red line), and 2019 (blue line), and the three layers: peat (gray), clay-sand (blue), and granite (brown). Brown ellipsoids represent the two granite boulders found to be distributed within the clay-sand volume of the 2018 excavation and the ten found in the 2019 excavation. The brown rectangle shows the location of the only ensemble of fractured rock found in the excavations.

The 2017 expedition excavated the volume that is defined by the black line in Figure 6, and found that the bottom was composed of compacted clay-sand mixture at a depth $0.6 \pm 0.1$ m. The compacted, post-liquefaction material was too hard to continue excavating by hand.

In 2018, a six-ton excavator was used in an attempt to reach the bedrock. The volume bounded by the red line in Figure 6 was excavated, with the sides sloping an average of 0.5:1, i.e. 0.5 m horizontal for every 1.0 m vertical, or ~27° from vertical, in accordance with local experience in this soil. At $4.7 \pm 0.1$-m depth, a grouping of fractured rock was discovered just northeast of the center line. After an hour of observing the stability of the sides, the principal investigator was cleared by the civil engineer safety officer to enter the pit. He scooped accumulated water into a bucket and found that the rock was closely packed shards of granite with dimensions varying between 0.02 m and 0.1 m.

The excavation had to be quickly abandoned because the sides of the water-saturated clay-sand mixture showed signs of fracture and sliding at various points down the slope. No samples were removed because we did not have time to do a careful and well-documented investigation. The dimension of the rocky bottom was at least the ~0.5 m of the cleared bottom, but the horizontal dimension of the rocky area could not be determined; it could be an extensive layer of fractured rock, fractured bedrock, or a localized deposit.

The 2019 expedition used two 14-ton excavators to dig the hole that is defined by the blue line shown in Figure 6. The slope of the sides averaged 1.5:1, i.e. 1.5 m horizontal for every 1.0 m vertical, or ~55° from vertical, to assure they would not collapse. Ten boulders were found throughout the excavation. Figure 7 shows two of these.

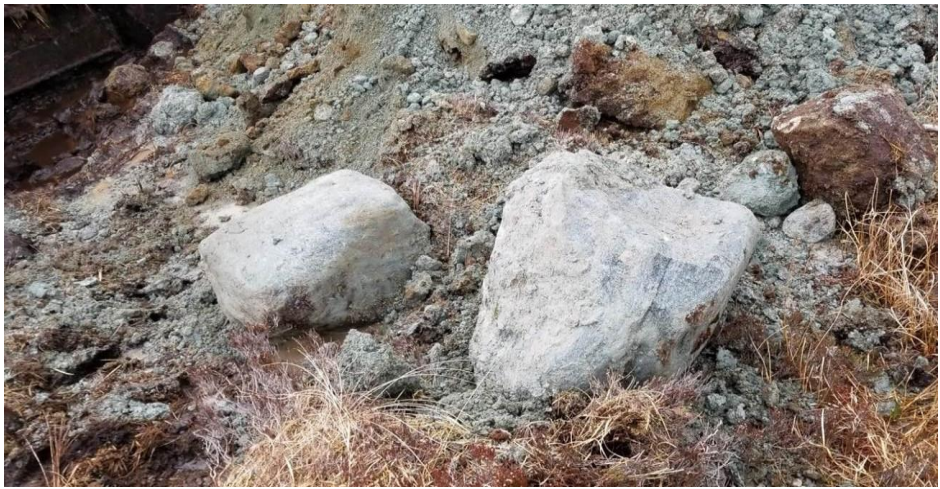

**Figure 7.** Two of the ten boulders found within the excavated volume are shown. Their diameters are approximately 0.4 m and 0.6 m, and they were distributed throughout the 633 m$^3$ of the 2019 excavation, but similar boulders were not observed on the surface of the peat bog.

Because the material above the rocky grouping of interest had been back-filled after the 2018 excavation, the precise positions of the boulders were not relevant to the 1985 event and they were not recorded by the excavator operators.

The operators were requested to excavate to the rocky layer at -4.7 ± 0.1 meters, stop, and alert the team. They did so; however, by the time they stopped and measured the depth, they had removed the volume of fractured rock in just one bucket load, demonstrating that it was a localized deposit, and then discharged it through the relocation process to a pile where it spread out. Although they showed us where that load lay, its relational context was lost. We encourage another group to re-excavate the site and look for fractured granite in the bedrock below our excavation; extreme care is recommended to preserve the context of fractured rock.

Water was pumped from the excavation. The muddy bottom was explored by hand. The rocks shown in Figure 8 were found between 5.0-m and 6.3-m depth. The smaller ones are consistent with those that were observed at 4.7 m in the 2018 excavation. The larger rocks have slightly rounded edges and they may not have been from that grouping.

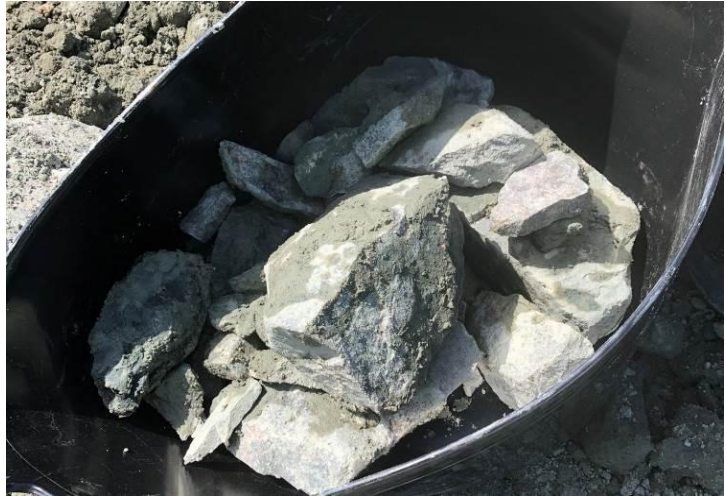

**Figure 8.** Granite shards from the volume below the grouping of fractured granite at 4.7 ± 0.1 m are shown. The rocks are covered with the fines from the clay-sand mixture which distorts their natural colors. Although rocks that are similar to the larger samples in Figure 8 were found on the surface of the peat, no collection of rocks similar to the single shattered boulder was found on the surface.

These granite rocks were examined with Energy Dispersive Spectroscopy [41] for evidence of large pressure gradients having altered the quartz in the granite. Streaks of darkened mineral was determined to be natural feldspar. No damage that was attributable to extreme pressures was found.

The excavation continued to a depth of 5.7 m, as illustrated by the rectangle outlined in blue in Figure 6. The west face of the crater, just west of the grouping of shards found in 2018, was washed with a pressure washer to better reveal its composition. Figure 9 shows a photo of the washed face.

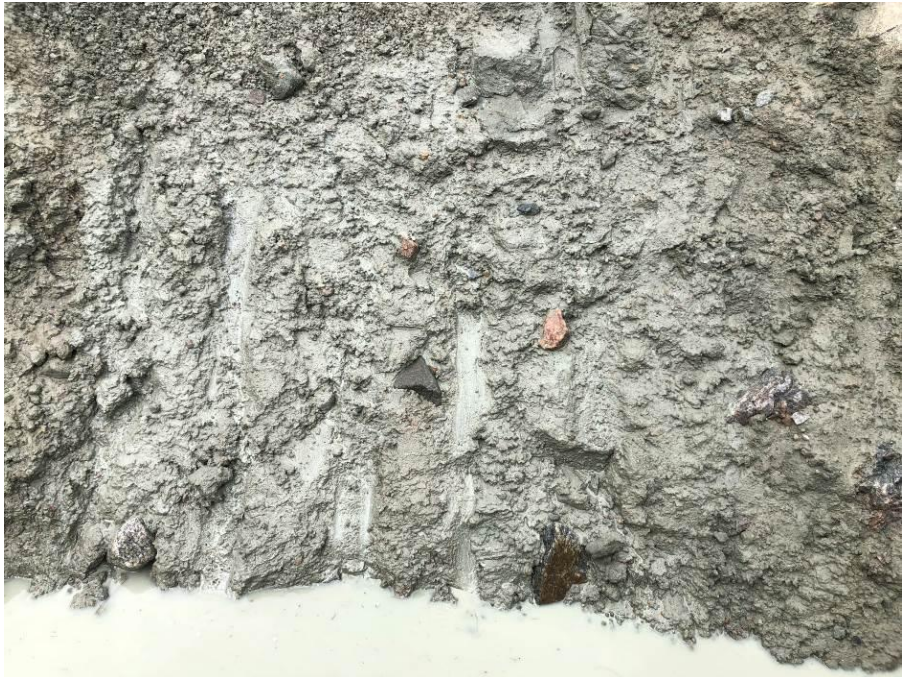

**Figure 9.** Pressure-washed face of the excavation's west side, adjacent to the grouping of shards at the 4.7 ± 0.1 m depth found in the 2018 excavation.

Figure 9 shows no evidence of a horizontal layer of shards or bedrock, demonstrating that the ensemble of fractured granite was an isolated one, approximately the size of a shattered boulder. This group of shards was in the projected path of the impactor and it was the only such grouping found in the excavation. Because boulders that are closer to the surface, but outside the projected path, were not shattered, we conclude that the ensemble of fractured granite was not shattered by pressure waves originating from energy deposited near the surface.

The decrease of pressure with increasing distance in the movies of the CTH simulations in Supplementary Material and the diameter of the fractured granite shown in Figure 3 indicate that a direct hit by an MQN may be required, and would be sufficient, to fracture a whole boulder. If so, and if the boulders were randomly distributed within the excavated volume, the probability of even one boulder being intercepted and fractured was only ~ 7%. Consequently, it is not surprising that only one collection of shards was found, and that it was well within the projected path of the penetrator. Therefore, we infer the hypervelocity object shattered the granite boulder after passing through 0.7 m of peat and 3.9 m water-saturated soft sediments.

We found at ~ 6.3-m depth, irregular boulders and large flat slabs of granite, with the vector normal to a slab inclined at ~30° to the vertical on the south, ~60° to the vertical on the north, and ~ 90 ° to the vertical in the middle, as shown in Figure 10. We did not find a uniform slab of bedrock that would have been a perfect witness plate of a quark-nugget passage by showing a cylinder of fractured granite extending into the earth. Broken slabs in disarray might be expected because our simulations give ~ 160 MJ/m (the equivalent of ~160 sticks of dynamite per meter) in granite to match the crater.

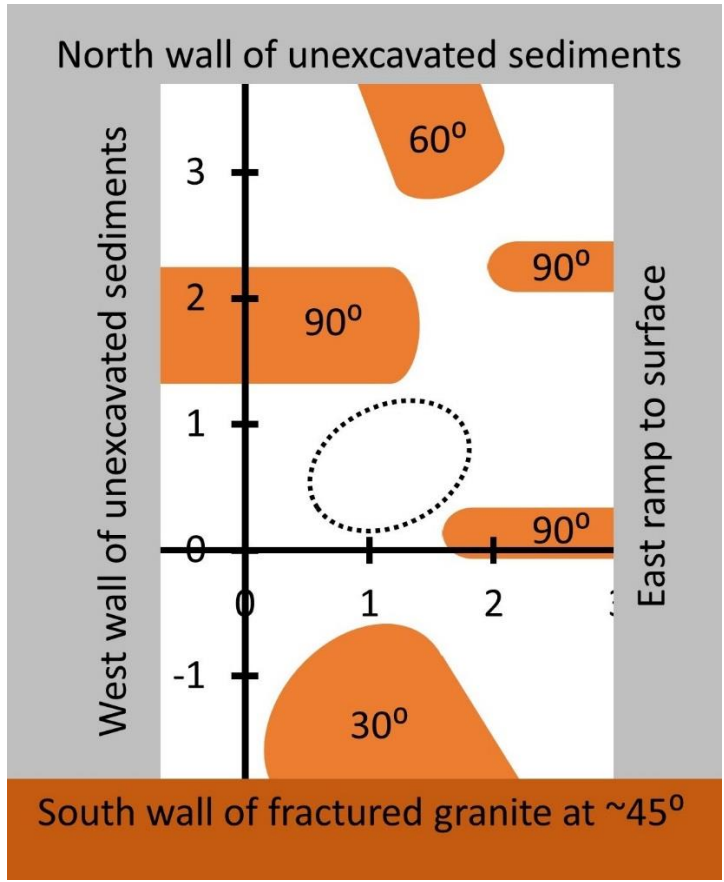

**Figure 10.** The layout of granite rocks (brown) and sediment walls (gray) at depth of ~ 6.3 to 6.5 m. The scale is in meters and the angles are between the vertical and the normal to the largest-area surface. The origin is directly below the center of the original impact crater on the surface with an estimated accuracy of +/- 0.3 m. The dotted ellipse shows the approximate projection of the shattered rock found in the 2018 excavation at 4.7-m depth to the 6.4-m depth shown here.

We could not determine whether the mixture of rocks and slabs at different angles to the horizontal, as shown in Figure 10, were characteristic of the site before the 1985 event or were caused by that event. Additional excavation directly beneath the grouping of fractured rock at ~4.7-m depth was blocked by large boulders or displaced slabs around that volume. These obstacles were too large to move with available equipment. In addition, the excavation from 4.8 m to 6.3 m had nearly vertical walls, which introduced a safety risk and precluded more excavation within the limitations of the project.

*3.6. Potential for Independent Validation of the 1985 Event*

The force equation for a high-velocity body with instantaneous radius $r_m$, mass $m$, and velocity $v$, moving through a fluid of density $\rho_p$ with a drag coefficient $K \approx 1$, is

$$F_e \approx K\pi r_m^2 \rho_p v^2. \tag{3}$$

MQNs have a velocity-dependent interaction radius [21] that is equal to the radius of their magnetopause

$$r_m \approx \left( \frac{2B_o^2 r_{QN}^6}{\mu_0 K \rho_p v^2} \right)^{\frac{1}{6}}, \quad (4)$$

in which $r_{QN}$ is the radius of the MQN of mass $m$ and mass density $\rho_{QN}$:

$$r_{QN} = \left( \frac{3m}{4\pi \rho_{QN}} \right)^{\frac{1}{3}}. \quad (5)$$

The interaction radius of an MQN varies as velocity $v^{-1/3}$ in Equation (4). Including that velocity dependence in the calculation with initial velocity $v_o$ gives velocity as a function of depth $x$ yields

$$v = v_o \left( 1 - \frac{x}{x_{max}} \right)^{\frac{3}{2}}, \quad (6)$$

in which $x_{max}$ is the stopping distance for an MQN:

$$x_{max} = \left( \frac{3m}{2\pi r_{QN}^2} \right) \left( \frac{\mu_o v_o^2}{2K^2 \rho_p^2 B_o^2} \right)^{\frac{1}{3}}. \quad (7)$$

The ~10 kg MQN that is inferred for the 1985 crater penetrates to $x_{max}$ = 3572 m for $\rho_p$ = 2020 kg/m$^3$.

Because we only explored the 1985 event to a depth of 6.5 m, it is possible for an independent team to re-excavate the site of the 1985 event to the bedrock and look for an extended volume of fractured granite. We marked the site to facilitate such an independent examination.

*3.7. Additional Limit on Bo to ≥ 4 × 10¹¹ T*

Comparing simulation results with observations from the crater implies that the crater was formed with 80 +/- 16 MJ/m energy deposition in the peat. The MQN mass that can deposit that energy density and the corresponding number of events per year were computed as a function of $B_o$ [15]. Table 1 summarizes the results to compare with observations.

**Table 1:** Maximum mass, mass capable of delivering 80 MJ/m and 1 kJ/m, and an estimate of the corresponding event rates for interstellar dark-matter mass density ~7 × 10$^{-22}$ kg/m$^3$ [1,6] and 250 km/s impact velocity, the relative velocity of the solar system through the dark matter halo. NA = Not Applicable and means that there was no solution.

| $B_o$ (10$^{12}$ T) | Max Mass (kg) | Mass of MQNs depositing ~80 MJ/m (kg) | Expected events/y on Earth capable of ~ 80 MJ/m | Mass of MQNs depositing >1 kJ/m (kg) | Expected events/y depositing > 1 kJ/m | Ratio of rate > 1 KJ to rate ~ 80 MJ/m |
|---|---|---|---|---|---|---|
| 0.1 | 7 × 10$^{-3}$ | NA | 0 | 7 × 10$^{-6}$ | 100,000,000 | NA |
| 0.2 | 2 × 10$^{-1}$ | NA | 0 | 2 × 10$^{-5}$ | 3,000,000 | NA |
| 0.3 | 3 | NA | 0 | 4 × 10$^{-5}$ | 800,000 | NA |
| 0.4 | 2 × 10$^{1}$ | 16.9 | 2,000 | 5 × 10$^{-5}$ | 600.000 | 350 |
| 0.5 | 1 × 10$^{2}$ | 13.9 | 3,000 | 8 × 10$^{-5}$ | 200,000 | 50 |
| 0.6 | 1 × 10$^{2}$ | 11.8 | 3,000 | 1 × 10$^{-4}$ | 50,000 | 20 |
| 0.7 | 7 × 10$^{2}$ | 10.3 | 900 | 1 × 10$^{-4}$ | 20,000 | 30 |
| 0.8 | 1 × 10$^{3}$ | 9.1 | 300 | 2 × 10$^{-4}$ | 20,000 | 50 |
| 0.9 | 7 × 10$^{3}$ | 8.2 | 20 | 2 × 10$^{-4}$ | 1,000 | 50 |
| 1.0 | 2 × 10$^{4}$ | 7.5 | 20 | 2 × 10$^{-4}$ | 1,000 | 60 |
| 1.1 | 4 × 10$^{4}$ | 6.9 | 20 | 2 × 10$^{-4}$ | 900 | 50 |
| 1.2 | 1 × 10$^{5}$ | 6.4 | 40 | 3 × 10$^{-4}$ | 200 | 60 |
| 1.3 | 1 × 10$^{5}$ | 5.9 | 20 | 3 × 10$^{-4}$ | 100 | 50 |
| 1.4 | 6 × 10$^{5}$ | 5.6 | 30 | 4 × 10$^{-4}$ | 100 | 40 |
| 1.5 | 7 × 10$^{5}$ | 5.2 | 20 | 4 × 10$^{-4}$ | 90 | 40 |
| 1.6 | 1 × 10$^{6}$ | 4.9 | 0.9 | 4 × 10$^{-4}$ | 20 | 30 |
| 1.7 | 3 × 10$^{6}$ | 4.7 | 0.4 | 5 × 10$^{-4}$ | 10 | 30 |
| 1.9 | 6 × 10$^{6}$ | 4.2 | 0.3 | 6 × 10$^{-4}$ | 10 | 30 |
| 2.0 | 9 × 10$^{6}$ | 4.0 | 0.2 | 6 × 10$^{-4}$ | 7 | 30 |
| 2.1 | 1 × 10$^{7}$ | 3.9 | 0.5 | 7 × 10$^{-4}$ | 2 | 30 |
| 2.3 | 5 × 10$^{7}$ | 3.6 | 0.06 | 8 × 10$^{-4}$ | 2 | 30 |
| 2.4 | 2 × 10$^{8}$ | 3.4 | 0.06 | 8 × 10$^{-4}$ | 0.4 | 10 |
| 2.6 | 2 × 10$^{8}$ | 3.2 | 0.003 | 9 × 10$^{-4}$ | 0.2 | 10 |
| 2.8 | 6 × 10$^{8}$ | 3.0 | 0.003 | 1 × 10$^{-3}$ | 0.2 | 10 |
| 3.0 | 1 × 10$^{9}$ | 2.8 | 0.005 | 1 × 10$^{-3}$ | 0.03 | 10 |
| 3.1 | 2 × 10$^{9}$ | 2.7 | 0.0003 | 1 × 10$^{-3}$ | 0.02 | 10 |
| 10.0 | 8 × 10$^{14}$ | 1.0 | 3 × 10$^{-8}$ | 7 × 10$^{-3}$ | 2 × 10$^{-7}$ | 10 |

Table 1 shows the MQN mass that is necessary to deposit 80 MJ/m in water-saturated peat as a function of $B_o$. We exclude $B_o < 4 \times 10^{11}$ T because the maximum MQN mass in distributions with $B_o < 4 \times 10^{11}$ T cannot deliver that energy deposition.

The last column shown in Table 1 gives the ratio of event rate with enough energy deposition (~1 kJ/m) to leave some geophysical evidence to the event rate that is sufficient for producing the crater in 1985 (~80 MJ/m). The ratio varies from 10 to 350 for the allowed range of $B_o$, and it indicates that there could be a sufficient number of events to study, if they can be identified and if access to the sites can be obtained.

*3.8. Event Rate of Non-Meteorite Cratering Events and Duplicative Constraint on Bo*

In addition to comparisons by MQN mass, an observed event rate can be compared to theoretical predictions of the event rate in Table 1. Three non-meteorite events in three years were cited in the Introduction. The estimated energy/meter deposited from Figure 3 above for the 2016 event [33] in Tamil, India that killed a man was ~80 MJ/m, which is comparable to the energy deposition in the County Donegal event. The 2015 event [32] in Rhode Island, USA, is consistent with ~1 kJ/m energy deposition. The 2014 event [31] in Managua, Nicaragua, is consistent with ~30 GJ/m deposited in soft sediment. The approximately annual event rate can be associated with MQN impacts delivering ≥1 kJ/m. Table 1 summarizes the event rate for that mass range as a function of $B_o$. An annual event rate appears to exclude $B_o > 2.3 \times 10^{12}$ T.

However, we need to interpret these results cautiously. We do not know what fraction of all events are observed and reported. If that fraction is small, then some values of $B_o \leq 2.3 \times 10^{12}$ T would also be excluded. On the other hand, MQNs can certainly survive transit through a portion of the solar chromosphere and photosphere and be decelerated (by the MQN magnetopause interaction with solar plasma) to less than the escape velocity from the solar system. A very small fraction of these trapped MQNs can receive sufficient angular momentum, by subsequent interaction with a planet, so that they are not absorbed into the sun. In principle, these captured MQNs can accumulate and enhance the dark-matter density inside the solar system, as compared to that of interstellar space. Our preliminary estimates provide an enhancement factor of ~300. Until adequate simulations of this aerocapture process are completed, we refrain from excluding $B_o$ values that are compatible with an enhancement of 300. Therefore, the upper excluded value that is based on event rate remains $B_o > 2 \times 10^{12}$ T and it is less restrictive than the constraint that is based on MQN mass in Section 3.7.

**4. Discussion**

*4.1. Consistency with MQN Impact*

Five points of agreement between theory, as interpreted by the simulations, and data from the three witness-plate layers combine to provide the second of many needed observations that are consistent with MQN dark matter.

1. The 4-m diameter crater is consistent with CTH simulations of ~80 MJ/m energy deposition. That energy/length is consistent with a 10 +/- 7 kg MQN with $4 \times 10^{11}$ T $\leq B_o \leq 3 \times 10^{12}$ T. It is not consistent with a meteorite, because no meteorite material was found and because the crater diameter is much too small to be within the range of meteorite craters. Meteorites must be either very aerodynamically shaped, which is very unlikely, or be at least ~

20-m diameter to be large enough to survive the transit through the atmosphere and create an impact crater. Meteorite craters are typically an order of magnitude larger in diameter than the meteorites that make them, so meteorites are usually found in craters with diameter ~200 m or larger. The smallest diameter crater associated with a meteorite in the last century impacted in 2007 at Caranacas, Peru. It was 13.5-m in diameter. The crater was at an altitude of 3,500 m. Its small diameter may be attributed to its not having to survive the densest part of the atmosphere. Non-meteorite craters are reported in the press approximately annually and are less than 12-m diameter, as noted in the Introduction. The lack of overlapping size and event rate suggests that craters, like the one studied in this paper, must be caused by a phenomenon other than a meteorite.

2. The CTH simulations show that the crater sides are formed by shear-planes and are smooth, as independently reported by the two Rangers investigating the day after the event. We found that smooth sides are in stark contrast to the irregular sides of craters that are produced by large explosives on the surface of the peat bog, so smooth crater sides are a distinguishing point of comparison.

3. The CTH simulations show that chunks of ejecta have sufficient velocity to be thrown clear of the site. Rangers reported the ejecta landed ≥10 m from the crater. The photograph presented in Figure 5 shows no ejecta near the crater, which confirms their report.

4. The "pointy depression" at the center of the crater bottom is consistent with the computed channel through the soft sediments and subsequent flow of material. Because the water-saturated soft sediments below the peat met all of the requirements for liquefaction [39] by the impact, the soft sediments must have liquefied and flowed back into the void to refill the central channel that is shown in Figure 3. Refilling would have occurred from the bottom, where the pressure from the overburden is the greatest, to the top. When the overburden pressure becomes too small to overcome viscosity, a "pointy" depression should remain, as observed on the day following the impact.

5. A volume of shattered granite was only found at 4.7-m depth and within the projected impact trajectory at 15° from vertical. All 10 boulders that were found outside the trajectory were intact. The uniqueness of the shattered granite and its location indicates the hypervelocity body that caused the crater in the peat layer also shattered the granite boulder at 4.7-m depth. Passage through the 0.7-m peat layer and 4.0-m soft-sediment layer with sufficient residual velocity to shatter the granite requires material strength that is much greater than the material strength of normal matter. The electromagnetic force holds normal matter together. The strong nuclear force is the only alternative. It holds quark nuggets together. Strong-force material strength, the corresponding nuclear mass density, and energy deposition in the MJ/m range in solid density matter are uniquely consistent with quark nuggets. Therefore, hypervelocity penetration through many meters or kilometers of solid or liquid density normal matter and energy deposition in the MJ/m range are a unique signature of an MQN. Therefore, the deeply buried and shattered granite is consistent with an MQN impact.

*4.2. Probability of Fractured Granite Attributable to the Impactor that Made the 1985 Crater*

The fifth point of comparison in Section 4.1 assumes that the fractured-granite deposit was caused by the impactor that produced the crater. We only found one fractured-granite deposit in the 633 m$^3$ excavation. It was at 4.7-m depth and within the calculated trajectory of the crater-forming impactor. The state of subsurface rock before the 1985 impact is uncertain, as with all non-meteorite impacts. Consequently, we cannot be certain that the highly localized and uniquely fractured granite was not fractured before the impact. Its association with the impact is only based on its location and uniqueness.

The null hypothesis is that the fractured granite at 4.7-m depth and the crater on the surface were not caused by the same event. If the probability of the null hypothesis is < 0.05, then the results are usually considered to be worthy of further investigation as possible evidence for a new phenomenon. We estimate the probability $P_{null}$ that the null hypothesis is true. $P_{null}$ has two factors:

1. $P_1$ = probability of the single shattered boulder being randomly located within the effective range $R_{eff}$ of the impactor trajectory for fracturing granite. $P_1$ = volume_ratio = $\pi R_{eff}^2 L/(633\ m^3$ volume of excavation), where L ~ 5m depth of excavation

2. $P_2$ = probability of 10 intact boulders being outside the effective range $R_{eff}$. Because the probability that one boulder is outside $R_{eff}$ is 1 - the probability it is inside $R_{eff}$ and since all 10 are assumed to be independently located, $P_2 = (1 - $ volume_ratio$)^{10}$.

Therefore, $P_{null} = (0.025\ R_{eff}^2) \times (1 - 0.025\ R_{eff}^2)^{10}$.

Figure 10 shows that the impactor trajectory is within one meter of granite slabs that are still intact, so $R_{eff} \leq 1$ m if the energy deposition in the clay is effective in fracturing granite. Granite fractures from the tensile stress after compression waves that originate inside the granite reflect off of the interface with lower-impedance media, as summarized in Section 3.1 from references [37,38]. If fracturing into shards requires energy being directly deposited inside the granite, then $R_{eff} \leq 0.45$ m, the boulders' mean diameter. The two estimates of $R_{eff}$ give probability $P_{null}$ between 0.005 and 0.02. Because we did not measure the exact location of each boulder as it was excavated, then the $P_2$ term is less certain but it is not sensitive to this number. Setting the less certain $P_2$ term to 1 still gives $P_{null}$ between 0.005 and 0.025. The probability that the fractured boulder is associated with the impact is 1.0 − $P_{null}$ is >98%. The high probability of association supports the consistency of the 1985 event with an MQN impact.

*4.3. Less than 20-m diameter craters are incompatible with normal-matter impactors.*

The 80 MJ/m that was deposited in the 1985 event requires the impactor to have been a hypervelocity body, in which the material strength is much less than the internal stresses. Hydrodynamic simulations [28] of the disintegration of large (1 m to 1 km in size) meteoroids in Earth's atmosphere show aerodynamic force, which is proportional to atmospheric density times the square of the velocity, causes it to decelerate, and produces a strong shock wave in front of it. The interaction compresses, heats, and ionizes atmospheric gas. Plasma temperatures can reach 25,000–30,000 K. The associated thermal radiation is absorbed by the surface material of the impactor and causes rapid ablation and vaporization. Rayleigh–Taylor instabilities strongly deform the body, which first breaks up in the center and then completely breaks into many small fragments that quickly slow to subsonic terminal velocity incapable of making a crater. The results are validated by comparison of the predicted light signatures with satellite-based observations and they are consistent with meteorites distributed over a wide area without a crater, as observed in Antarctica.

If the meteor is sufficiently large and sufficiently aerodynamic, then it reaches the ground intact with a significant fraction of its mass and it is still moving at hypervelocity speeds. It then produces an impact crater that is accompanied by meteorite material at the impact site. The dynamics that are associated with passage through Earth's atmosphere assure zero to a very small fraction of meteors with < 20-m diameter survive and maintain sufficient speed to cause an impact crater [28]. Because the diameter of an impact crater is typically ~ 5% the diameter of the impactor, impact craters that are less than ~100 m in diameter are inconsistent with normal-matter impactors. Therefore, the 3.5-m diameter crater from the 1985 event is inconsistent with the normal-matter impactor.

*4.4. Normal-matter impactors delivering the inferred energy to the peat are incompatible with shattering granite 4.7 m below the surface.*

The impactor in the 1985 event delivered ~80 MJ/m to the 0.7 m of peat, penetrated 4.0 m of water-saturated soft sediments, and still had enough momentum to shatter the granite boulder with an observed diameter of ≥0.6 m. Any MQN that deposits ~80 MJ/m in the peat will also deposit, proportional to its mass density, ~160 MJ/m in the granite. This is well in excess of the ~1 MJ/m required in granite to shatter the boulder, as shown in Figure 3.

Transit through solid or liquid density media would require surviving dynamic forces more than 1000 times those that were experienced in the atmosphere and discussed in Section 4.3, so such transits are prohibited for normal matter. In addition, conservation of linear momentum assures an approximately spherical body (not a long rod penetrator) is decelerated with an e-folding distance of approximately its diameter times the ratio of impactor density to media density. Meteorites typically make a crater approximately 20 times the meteor diameter. Even if the impactor is not vaporized upon impact, a normal-matter impactor would lose most of its velocity within a few tenths of the crater diameter. Consequently, we can rule out normal-matter impactors as the cause of shattered granite boulders that are 4.7 m below the 3.5-m diameter crater in Ireland.

However, the strong nuclear force determines the material strength of an MQN. They are indestructible in interactions at even 250 km/s. The corresponding mass density is nuclear density $> 7 \times 10^{17}$ kg/m$^3$ and this assures that their momentum will let them penetrate many meters or even kilometers into Earth, as discussed in Section 3.6.

*4.5. Alternative Explanations*

Quark nuggets, neutronium, and black holes have mass densities that are greater than the required value. However, neutronium is not stable outside of neutron stars, and black holes are small enough to provide the local density of dark matter and provide at least one impact per year reported in the press, i.e. ~10 kg mass, would have evaporated in about 150 y, which is much shorter than the time over which the effects of dark matter have been stable. Therefore, crater formation by quark-nugget impact is the only explanation that we have found that fits the data and is consistent with established physics.

These results are also consistent with any other hypervelocity nuclear-density impactor of mass ~ 5 kg and interaction physics that are capable of depositing ~ 80 MJ/m. Axion Quark Nuggets (AQNs) [11] have been proposed as an extension to the Standard Model. Their predicted characteristics also satisfy these requirements. Therefore, the results that are reported in this paper also support the AQN candidate for dark matter and may be used to test other dark-matter hypotheses.

*4.6. Limitations to Evidence and Need for Systematic Study*

The 1985 impact is only the second reported event that is consistent with the MQN hypothesis. Many more are needed to conclude that MQNs exist. The three non-meteorite events cited in the Introduction could be investigated as additional quark-nugget events. Additional candidates are listed in the Supplementary Results: Additional candidate sites for MQN impacts in County Donegal. However, the dates of these potential events are unknown, and there may be competing processes for producing crater-like holes in otherwise flat peat bogs. Gaining physical and administrative access to investigate these additional sites may be difficult. Our investigation of the 1985 impact in Ireland required fifteen years, even with a supportive local community and national authority.

Additional and independent excavation of the 1985 event in County Donegal is lower risk and it could independently confirm or invalidate our result by determining if the bedrock shows the expected cylindrical hole of fractured granite with radius of fracture decreasing with increasing depth. In addition, the expedition could determine whether the tilted granite slabs and granite rocks at 6.3-m depth are a universal feature of the bedrock in the area or were caused by the 1985 event. The latter case would provide additional evidence of large and local energy deposition at depth. The information in Appendix A should be helpful to such an expedition.

A systematic study is necessary, even with additional evidence from non-meteorite craters. Obtaining the results presented in this paper required fifteen years, including the time to obtain permission to excavate from supportive land owners and Irish national authorities. Although such events apparently occur annually on Earth, obtaining permission and excavating each one is impractical. If remote acoustic sensing [42] of the subsurface interface between granite bedrock and soft sediments could be further developed to provide a profile on the interface with ~10-cm resolution through ~ 10 m of soft sediments, then the pattern that is shown in Figure 10 might be identified as uniquely associated with the impact event. If so, new events could be explored if access to the site can be secured. In addition, the rest of the peat bog in County Donegal could be non-destructively mapped to identify additional sites that occurred over the last 3500 years. Appendix B presents a list of candidate sites for MQN impacts in County Donegal. With this method, a statistically significant set of data might be obtainable.

Whether or not such a technology can be developed, the event that is reported in this paper motivates developing and deploying a constellation of three satellites at 51,000 km altitude to look for RF signatures of MQNs after they transit the magnetosphere [34]. Such a space-based system would provide a real-time search for MQNs based on their predicted Doppler-shifted-radiofrequency signature and it is the best approach for the necessary and systematic study of the MQN hypothesis for dark matter.

**5. Conclusions**

We report computer simulations of the MQN energy deposition in water-saturated peat, soft sediments, and granite, and report the results from excavating such a crater. The >98% probability that the fractured boulder is associated with the impact (Section 4.2) and the five points of agreement between the simulation results and the observations (Section 4.1) support the inference that the 1985 event is consistent with an MQN impact. This is the second event found to be consistent with MQNs. The first is described in reference [17]. However, many additional non-meteorite impacts with a similar effect on deeply buried rock and/or additional tests that stress different aspects of the MQN hypothesis are needed in order to conclude whether or not MQNs exist and contribute to dark matter. The results also redundantly constrain $B_o$ to $\geq 4 \times 10^{11}$ T, which is consistent with the previously published most likely values of $B_o = 1.65 \times 10^{12}$ T +/- 21%.

Although these results are consistent with MQNs, they are also consistent with any other hypervelocity nuclear-density impactor of mass ~ 5 kg and interaction physics that are capable of depositing ~ 80 MJ/m, such as Axion Quark Nuggets. The results may be also be consistent with some other phenomenon unknown to us. If such candidates are found, they may also be viable candidates for dark matter.

Non-meteorite craters are reported in the popular press approximately once per year. That frequency of reported events suggests a much larger event rate that may offer an opportunity to test hypotheses for dark matter.

**Supplementary Materials:** Movies of pressure, mass density, and temperature from the CTH simulations available at VanDevender, J. Pace; Schmitt, Robert; Simulations of magnetized quark nugget dark matter in three-layer witness plate, Dryad, Dataset, 2020, https://doi.org/10.5061/dryad.cc2fqz641.

**Author Contributions:** Conceptualization, Aaron VanDevender; Data curation, J. Pace VanDevender and Niall McGinley; Formal analysis, J. Pace VanDevender and Robert G. Schmitt; Funding acquisition, J. Pace VanDevender; Investigation, J. Pace VanDevender, Robert G. Schmitt, Niall McGinley, David G. Duggan, Seamus McGinty, Aaron VanDevender, Peter Wilson, Deborah Dixon, Helen Girard and Jacquelyn McRae; Methodology, J. Pace VanDevender, Robert G. Schmitt, Niall McGinley, Aaron VanDevender and Peter Wilson; Project administration, J. Pace VanDevender; Resources, J. Pace VanDevender; Software, Robert G. Schmitt; Supervision, J. Pace VanDevender; Validation, Robert G. Schmitt and Aaron VanDevender; Visualization, Robert G. Schmitt; Writing – review & editing, Robert G. Schmitt, Niall McGinley, David G. Duggan, Seamus McGinty, Aaron VanDevender, Peter Wilson, Deborah Dixon and Helen Girard.

All authors have read and agreed to the published version of the manuscript.

**Funding:** This research was primarily funded by VanDevender Enterprises and volunteers. In addition, New Mexico Small Business Assistance to Sandia National Laboratories funded the CTH hydrodynamic simulations.

**Institutional Review Board Statement:** Not applicable

**Informed Consent Statement:** Not applicable

**Data Availability Statement:** All final analyzed data generated during this study are included in this published article and associated supplementary information.

**Acknowledgments:** We gratefully acknowledge S. V. Greene for first suggesting we consider quark nuggets and Mark Boslough for critical review and suggestions. Ranger David Dugan provided greatly appreciated guidance and assistance in obtaining the necessary permit from the National Parks and Wildlife Service. Mr. Charlie Callahan of Glendowan provided essential guidance and assistance from the point of view of local residents with rights to use this land. Mr. Cathal Moy and his associates were invaluable; without their knowledge of local soil conditions and exceptional skill with operating their excavators, the project could not have been accomplished. Volunteers Rebecca Stair and Nathan Girard assisted in the 2018 and 2019 excavations respectively. Finally, Jesse A. Rosen edited this paper and improved the structure, logic, and flow.

This work was supported by VanDevender Enterprises, LLC. The CTH simulations were supported by the New Mexico Small Business Assistance (NMSBA) Program through Sandia National Laboratories, a multi-program laboratory operated by National Technology and Engineering Solutions of Sandia (NTESS), a wholly owned subsidiary of Honeywell International, with support from Northrup Grumman, Universities Research Association and others., for the U.S. Department of Energy's National Nuclear Security Administration. By policy, work performed by Sandia National Laboratories for the private sector does not constitute endorsement of any commercial product.

**Conflicts of Interest:** The authors declare no conflict of interest.

## Appendix A. Excavations

To assist another team to extend the excavation into the bedrock and independently test and extend our findings, the three excavations are described in this section. Please check the Acknowledgements for the names of essential team members from County Donegal.

The 2017 expedition cleared out debris and plant growth by hand. The bottom, at depth $-0.6 \pm 0.1$ m, was compacted clay-sand mixture.

The 2018 expedition employed a single Hitachi EX-60, 6-ton excavator shown in Figure A1 with the 4-m diameter crater drained by the channel on the right edge. The site was excavated with the sides sloping at 27° to the vertical, in accord with local experience in this soil. However, the excavation had to be quickly abandoned because the sides of the water-saturated clay-sand mixture showed signs of fracture and sliding at various points down the 27° slope.

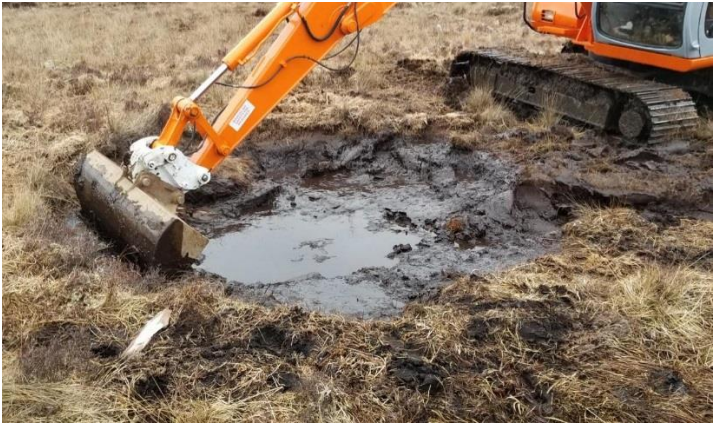

**Figure A1.** View of the 4-m diameter crater from 1985, drained with a channel shown to the down sloping terrain on the right.

The 2019 expedition employed two Doosan 140LC, 14-ton excavators. One was on a ramp inside the excavation and moving material to the surface. The second excavator relocated each scoop of material to a safe distance from the hole to avoid increasing pressure on the soil adjacent to the hole and provide a flat surface for the second excavator to traverse. The slope of the sides was approximately 55° from vertical as shown in Figure A2. That slope held.

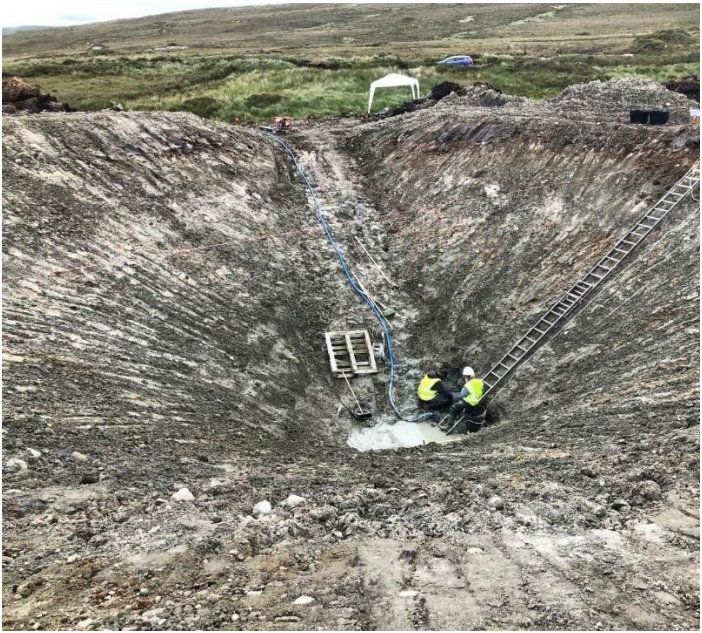

**Figure A2.** Excavation in 2019 to depth of 5.7 m and showing access ramp, submersible water pump and hose, and exit ladder. Examining the bottom and recovering rock fragments had to be done by feel at this stage.

An additional 1.5 m of material, plus a water-collecting hole for the submersible pump, was excavated to look for bedrock. From 4.8-m to 6.3-m depth, we found irregular boulders, smaller rocks, and large flat slabs of granite with their normal vector inclined at 30° to the vertical on the south side, and 60° to the vertical on the north, as shown in Figure A3.

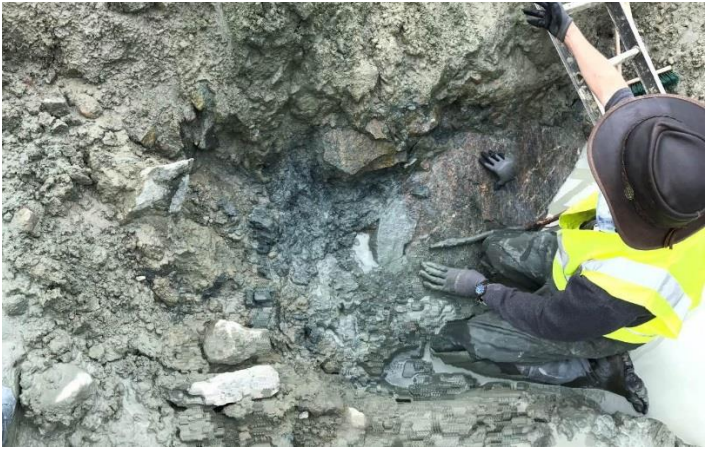

**Figure A3.** Photo of the collection of granite rocks found at approximately 6.3-m depth to the southeast of the center of the crater. The normal to the slab on the right is inclined at $30^0$ to the vertical.

We did not find a uniform slab of bedrock and could not determine if the mixture of rocks and slabs at different angles to the horizontal were characteristic of the site before the 1985 event or were caused by that event. Additional excavation directly beneath the grouping of fractured rock at ~4.5-m to ~4.8-m depth was blocked by two large boulders or displaced slabs to either side of that volume. These obstacles were too large to move with available equipment. In addition, the excavation from 4.8 m to 6.3 m had nearly vertical walls, which introduced a safety risk that precluded more excavation within the limitations of the project.

If another group re-excavates the site to examine the bedrock and search for the signature of MQN passage, i.e. a cylinder of fractured granite extending well into the earth, extreme care is recommended below the 6.3-m depth to preserve the context of fractured rock.

In accord with our permit, the site was first filled with the rock and clay-sand mixture and topped with the peat layer. A wooden pole was driven into the peat to mark the center of the original 1985 impact crater. Three orange plastic stakes are located on elevated mounds at 1) 21.9 m to the south, 2) 26.76 m to the west, and 3) at 40.12 m at 61.5° north of west. Surveyor's lines from each stake connect the stake to the center post in hopes that another expedition could easily find and re-excavate the site.

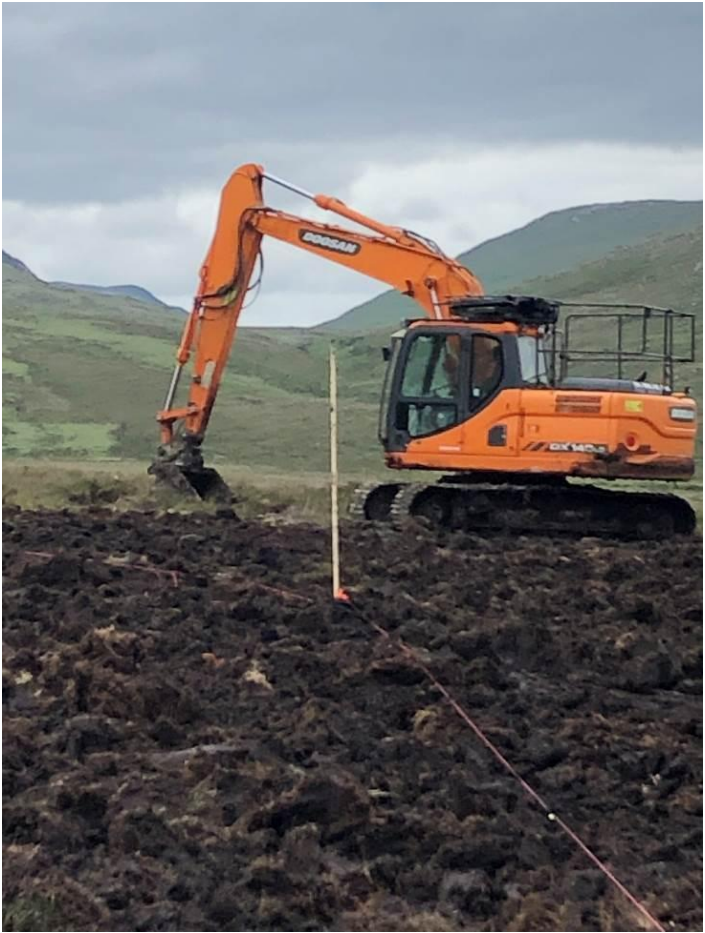

**Figure A4.** Wooden stake and three survey lines mark the center of the original impact crater for subsequent expeditions to extend the investigation into the granite bedrock below 6.5-m depth.

**Appendix B: Additional candidate sites for MQN impacts in County Donegal**

In 2006, we conducted an aerial survey of about 600 km2 of peat bog to look for other evidence of deformations from massive objects. We found at least two additional holes and inspected them on the ground: 1) approximately 4-m by 5-m diameter and 2 m deep at 54° 55.362' N and 8° 15.260' W and 2) approximately 4.8-m by 5.1-m diameter and 2.1 m deep at 54° 55.434' N and 8° 15.002' W. Since no one reported witnessing their being formed, we could not confirm that they were associated with impacts.

Since the aerial search in 2006, the resolution in the Google maps covering the western portion of the peat bog has been improved to the point that the maps are useful for a survey. Water flowing below the peat can create multiple holes aligned along the flow in peat bogs. Other mechanisms may also produce holes. Therefore, a survey of isolated round holes, like the 1985 event but without eye witnesses, will only give an upper limit to the event rate. A survey that was informed by the examination of the two holes found in 2006 was conducted in 2014. The survey consisted of 200 randomly selected areas in a square defined by the GPS coordinates of the opposing corners (54.918855, -8.222008) and (54.977614, -8.421822). The chosen area had adequate resolution and did not include any human structures. It was a peat bog with reeds growing on top of the older peat. The total area surveyed in the 200 samples was 3 km2. The survey identified 33 circular depressions like the two we qualified in the ground-based survey. The 33 positions are shown in Table A1.

**Table A1: Coordinates and description of deformations in peat-bog survey**

| GPS Coordinates | Diameter | Description |
|---|---|---|
| 54.920869,-8.398206 | 6m ± 2m | very circular, more faded than other shapes |
| 54.921385,-8.296381 | 6m ± 2m | circular, lighter strip running through circle |
| 54.926524,-8.310256 | 6m ± 2m | circular, faded, extension from bottom right of circle |
| 54.927851, -8.34372 | 4m ± 2m | very circular, more faded than other shapes |
| 54.929398,-8.261827 | 3m ± 2m | circular, nodule on top right and left of circle |
| 54.931636,-8.270241 | 2m ± 1m | circular, little nodule on top right of circle |
| 54.931716,-8.225976 | 4m ± 2m | circular, dip on bottom right of circle |
| 54.93359,-8.355017 | 4m ± 2m | circular, tiny dip at top left corner, surrounded by white |
| 54.935795,-8.269017 | 3m ± 2m | circular, sort of flat on top and bottom |
| 54.93674,-8.265257 | 2m ± 1m | very circular, surrounded by white |
| 54.937784,-8.267118 | 3m ± 2m | circular, nodule on top left of circle |
| 54.938974,-8.274889 | 4m ± 2m | circular, dip on top left of circle |
| 54.939262,-8.267507 | 4m ± 2m | circular, diagonal oval shape |
| 54.940387,-8.322002 | 3m ± 2m | very circular, more faded than other shapes |
| 54.942222,-8.319636 | 4m ± 2m | circular, white in center of circle |
| 54.943955,-8.276018 | 2m ± 1m | very circular, surrounded by white |
| 54.944405,-8.278464 | 2m ± 1m | very circular, surrounded by white |
| 54.946506,-8.292173 | 2m ± 1m | very circular, small |
| 54.947903,-8.23144 | 6m ± 2m | circular, two small rounded extensions at bottom |
| 54.949308,-8.399971 | 4m ± 2m | circular, nodule on bottom left of circle |
| 54.949494,-8.298809 | 5m ± 2m | circular, upright oval looking, faded |
| 54.951723,-8.30839 | 5m ± 2m | circular, diagonal oval shape |
| 54.95572,-8.265512 | 3m ± 2m | circular, slightly greater width than height |
| 54.956526,-8.24444 | 3m ± 2m | circular, small dip on bottom of circle |
| 54.962613,-8.273741 | 1m ± 0.5m | circular, tiny nodule on right side of circle |
| 54.964019,-8.305927 | 4m ± 2m | very circular, more faded than other shapes |
| 54.964546,-8.269087 | 2m ± 1m | circular, slightly greater height than width |
| 54.964943,-8.235828 | 2m ± 1m | very circular, more faded than other shapes |
| 54.965639,-8.237676 | 5m ± 2m | circular, nodule on bottom right of circle |
| 54.969051,-8.261765 | 3m ± 2m | circular, bit cut off bottom right of circle |
| 54.969271,-8.277284 | 3m ± 2m | circular, nodule on top right of circle |
| 54.970346,-8.377996 | 1m ± 0.5m | circular, diagonal oval shape |
| 54.971057,-8.320993 | 3m ± 2m | circular, extension from bottom of circle |

Poisson statistics gives a 95% confidence for an upper limit of 11 ± 3.7 events per km$^2$. Their diameters ranged from 2 ± 1 m to 9 ± 2 m. The crater from the 1985 event has changed little in 33 years and should last at least 100 years under the same environmental stresses. The extrema of 100 and 200 years for the time period give an estimated event rate of 0.1 to 0.05 km$^{-2}$ yr$^{-1}$. Since the area of the earth is ~$5 \times 10^8$ km$^2$, the corresponding global event rate is between $30 \times 10^6$ and $60 \times 10^6$ events per year. Such a large number of potential events illustrates the likelihood of other phenomena forming holes in peat bogs and the importance of eyewitnesses to impacts.